\documentclass{iopconfser}
\usepackage{siunitx}
\usepackage{graphicx}
\usepackage{subcaption}
\usepackage{cite}
\usepackage{upgreek}
\usepackage{hyperref}
\begin{document}


\title{First results from the E302 efficiency--instability experiment at the FACET-II facility}

\author{O~G~Finnerud$^{1}$,
E~Adli$^{1}$,
R~Ariniello$^{2}$,
S~Corde$^{3}$,
T~N~Dalichaouch$^{4}$,
C~Emma$^{2}$,
S~Gessner$^{2}$,
C~Hansel$^{5}$,
M~J~Hogan$^{2}$,
C~Joshi$^{6}$,
D~Kalvik$^{1}$,
A~Knetsch$^{2}$,
C~A~Lindstr{\o}m$^{1}$,
M~Litos$^{5}$,
N~Majernik$^{2}$,
K~A~Marsh$^{6}$,
B~D~O'Shea$^{2}$,
I~Rajkovic$^{2}$,
S Rego$^{3}$,
D~Storey$^{2}$
and
C~Zhang$^{6}$
}
\affil{$^{1}$Department of Physics, University of Oslo, 0316 Oslo, Norway}
\affil{$^{2}$SLAC National Accelerator Laboratory, Menlo Park, California 94025, USA}
\affil{$^{3}$Laboratoire d’Optique Appliquée (LOA), CNRS, École polytechnique, ENSTA, Institut Polytechnique de Paris, Palaiseau, France}
\affil{$^{4}$Department of Physics and Astronomy, University of California Los Angeles,
Los Angeles, California 90095, USA}
\affil{$^{5}$Department of Physics, Center for Integrated Plasma Studies, University of Colorado Boulder, Boulder, 
Colorado 80309, USA}
\affil{$^{6}$Department of Electrical and Computer Engineering,
University of California Los Angeles, Los Angeles, California 90095, USA}
\email{o.g.finnerud@fys.uio.no}
%


\begin{abstract}
The beam-breakup (BBU) instability in plasma accelerators is seeded by a transverse offset between the driver and a trailing bunch. The BBU instability induces oscillations in the trailing bunch, which are detrimental to its beam quality. When the instability is large, assuming little mitigation from ion motion and energy spread, the beam suffers emittance growth, and charge can be kicked transversely out of the plasma channel. The detrimental effect on beam quality is substantially worse at high efficiencies, which places constraints on the achievable power efficiency in applications such as linear colliders, where maintaining the beam quality is required. In this paper, we present the first experimental signatures of the BBU instability in data taken in the E302 experiment at the FACET-II facility at SLAC National Accelerator Laboratory. We use a specific beam-optical setup and a novel method to probe for transverse instabilities on diagnostic screens downstream of a magnetic dipole spectrometer. We complement the analysis with full 3D particle-in-cell (PIC) simulations of the plasma interaction using similar driver and trailing bunch parameters on a simulated FACET-II spectrometer.
\end{abstract}

\section{\label{sec:level1}Introduction}
Plasma accelerators~\cite{veksler, Tajima, beamdriven}, which replace RF cavities with plasma sources as the accelerating modules, have in recent years achieved several milestones on the path to becoming viable for use in linear colliders. Experimental results from laboratories such as SLAC and DESY have shown acceleration with high efficiency~\cite{Litos2014}, energy-spread preservation~\cite{efficienctpreserved}, and emittance preservation~\cite{Lindstr_m_2024}. 

Transverse stability of the electron beam during acceleration is needed to ensure emittance is conserved. At higher power-transfer efficiencies in a plasma accelerator, this becomes increasingly difficult due to strong transverse forces and resonance effects. The high charge of the trailing bunch needed to effectively beam load the longitudinal accelerating field causes large transverse wakefields \cite{STUPAKOV} if the bunch is misaligned. The driver-to-trailing-beam power transfer efficiency has been found in analytical studies~\cite{lebedev} as the key parameter predicting the strength of the BBU instability. The relation, named the efficiency--instability relation, is given as~\cite{lebedev}
\begin{equation}
    \eta_{t} = \frac{\eta_{p}^{2}}{4(1-\eta_{p})},
    \label{efficiencyinstability}
\end{equation}
with $\eta_{p}$ being the efficiency and $\eta_{t}$ being the strength of the instability normalised by the plasma-column focusing force. It is trivial to see that at high efficiencies, the instability becomes very large as the denominator approaches zero. Because of this scaling, measuring the BBU instability, which has not been done experimentally, is important. As it stands, the relation effectively constrains the efficiencies achievable in a plasma accelerator while maintaining sufficient beam quality for collider and free-electron laser applications. Furthermore, in a multi-stage accelerator, the efficiency could be further constrained as even a small instability would compound over a substantial total length of the accelerator. 

Recent studies using PIC codes \cite{Parametricmapping} have shown that the efficiency--instability relation above in fact describes a lower bound on the strength of the instability for a given efficiency:
\begin{equation}
    \eta_{t} \geq \frac{\eta_{p}^{2}}{4(1-\eta_{p})}.
    \label{efficiencyinstabilityineq}
\end{equation}
Additionally, Ref.~\cite{Parametricmapping}  shows that the strength of the instability can vary by orders of magnitude depending on the normalized wake radius and strength of the loaded accelerating field. As such, mapping the strength of the instability to parameters that can be adjusted in a plasma-accelerator facility is vitally important for future applications. In this paper, we present experimental results from the E302 experiment at the FACET-II facility, where the instability was probed using a novel method with the magnetic dipole spectrometer. This article discusses the first steps towards demonstrating the instability--efficiency relation.

\section{Method}
\begin{figure*}
    \centering
    \includegraphics[width = 1\linewidth]{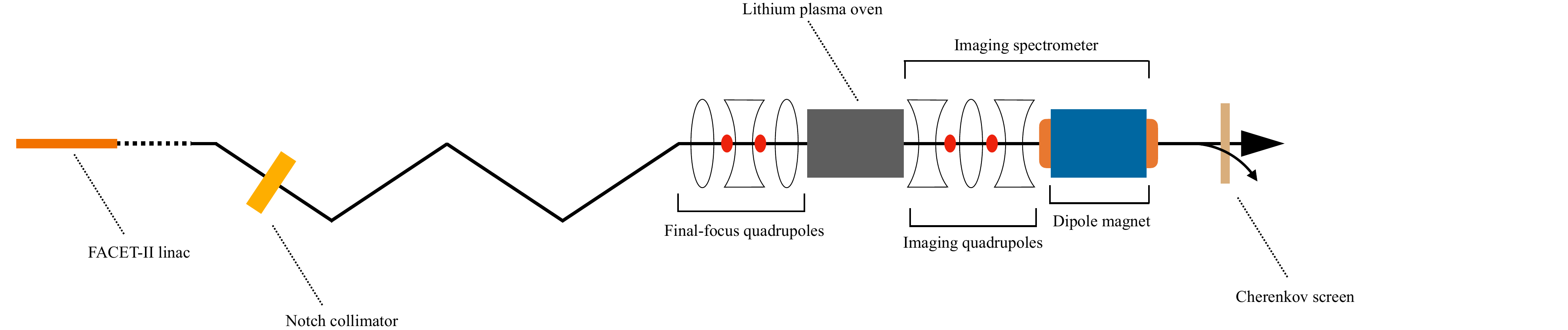}
    \caption{Schematic of the relevant parts of the beamline at the FACET-II facility for the E302 experiment (not to scale). The beam trajectory is indicated by the black arrow.}
    \label{Fig1}
\end{figure*}

The data presented was collected at the FACET-II facility at the SLAC National Accelerator Laboratory; a user facility that produces high-energy, low-emittance electron beams aimed at a lithium-vapour plasma source for plasma-wakefield experiments \cite{FACETfacility}. The elements in the beamline relevant for this paper are shown in Fig.~\ref{Fig1}. The main tool for analysing the transverse properties of an accelerated electron beam at the FACET-II facility is the dispersive magnetic Cherenkov-based spectrometer \cite{CHERENKOV}. After the beam exits the plasma, a magnetic quadrupole triplet focuses and magnifies it onto the diagnostic screens. Following the quadrupole triplet, a magnetic dipole disperses the beam in the vertical plane according to its energy distribution \cite{DOUGFACET}, as described in~\cite{ADLITRANSVERSE}. The transverse distribution of the beam on the diagnostic screen is a linear transformation of its width and divergence at the object plane. Typically, as in this analysis, the object plane is directly after the plasma. Therefore, with a known lattice (e.g., magnet strengths and element lengths), it is possible to back-propagate the spectrometer image to infer the transverse beam distribution exiting the plasma. 

The simplest and most common spectrometer setup is ``point-to-point imaging", where the transverse distribution on the screen at a specific energy is just a scalar multiplication of the distribution at the object plane. The disadvantage of this imaging setup is that one loses information about the angular distribution of the electron beam exiting the plasma. Furthermore, as one deviates from the imaging energy, the beam divergence begins to influence the distribution on the screen, leading to an unknown back-propagation from the screen to the object plane.

Another way to image the beam, developing the methods described in Ref.~\cite{ADLITRANSVERSE} further, is to measure its divergence while ignoring its spatial information. As the misaligned trailing bunch oscillates about the plasma axis, it behaves as a harmonic oscillator driven by coupling to the electron sheath. Therefore, the transverse angle of the longitudinal slices will also oscillate. As opposed to the spatial distribution of the bunch, the angular distribution can be accurately estimated across energy slices if the spectrometer is set up correctly. The setup requires the imaging energy to be either significantly lower or higher than the energy of the accelerated bunch. The bunch will then be located at energy slices that have a significant deviation from the imaging energy, and for those slices, the bunch's angle and divergence dominate its position on the screen, as shown in Ref.~\cite{ADLITRANSVERSE}. Since the magnetic lattice is known, one can then convert the spectrometer image to an angle--energy distribution as shown in Ref.~\cite{E302planning}.

\begin{figure}
    \centering
    \includegraphics[width = 0.45\linewidth]{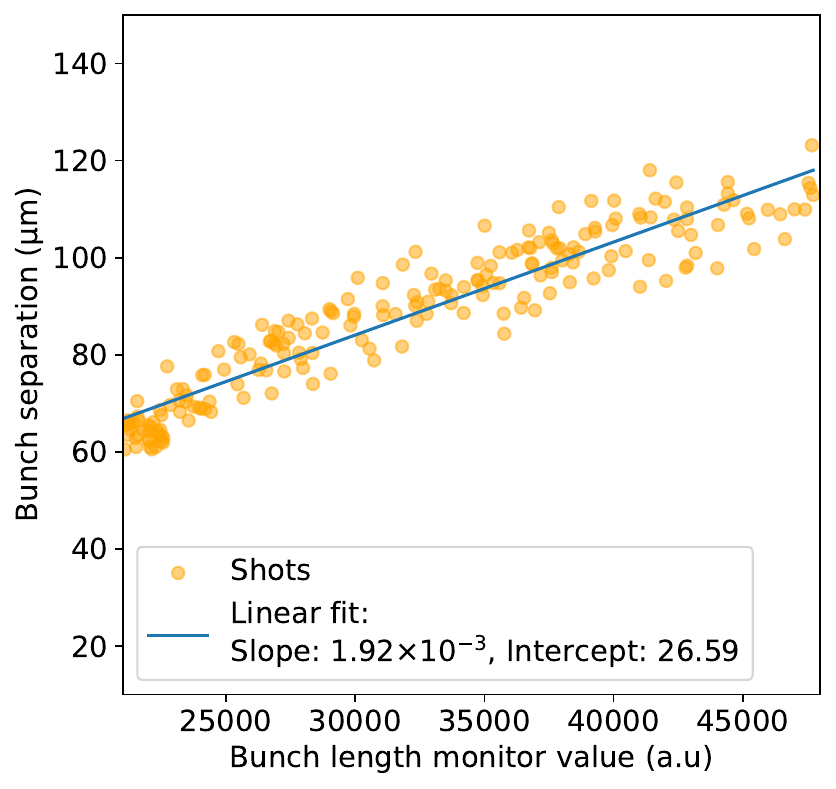}
    \caption{The correlation between bunch spacing found from the electro-sampling crystal and the readback value from the bunch length monitor in BC14 obtained from a L2 linac section phase scan dataset.}
    \label{Fig2}
\end{figure}

In the data analyzed, we scanned the phase of the L2 linac section at FACET-II, which induces an energy correlation with the longitudinal position within the beam. By using collimators in a dispersive section of the beamline, we separate the single beam in energy, and hence longitudinally due to the energy correlation, into two bunches. The bunch separation is then expected to correlate well with the L2 phase. The bunch separation was measured on a shot-to-shot basis using the electro-optic sampling setup at FACET-II \cite{EOS}. During phase adjustment, the correlation between the bunch length monitor signal in bunch compressor 14 (BC14) and the bunch separation of the driver and trailing bunch was approximately linear, as shown in Fig.~\ref{Fig2}. This correlation is used throughout the analysis to estimate the bunch separation for the experimental data.

\begin{figure}[h!]
    \centering
    \includegraphics[width = 0.45\linewidth]{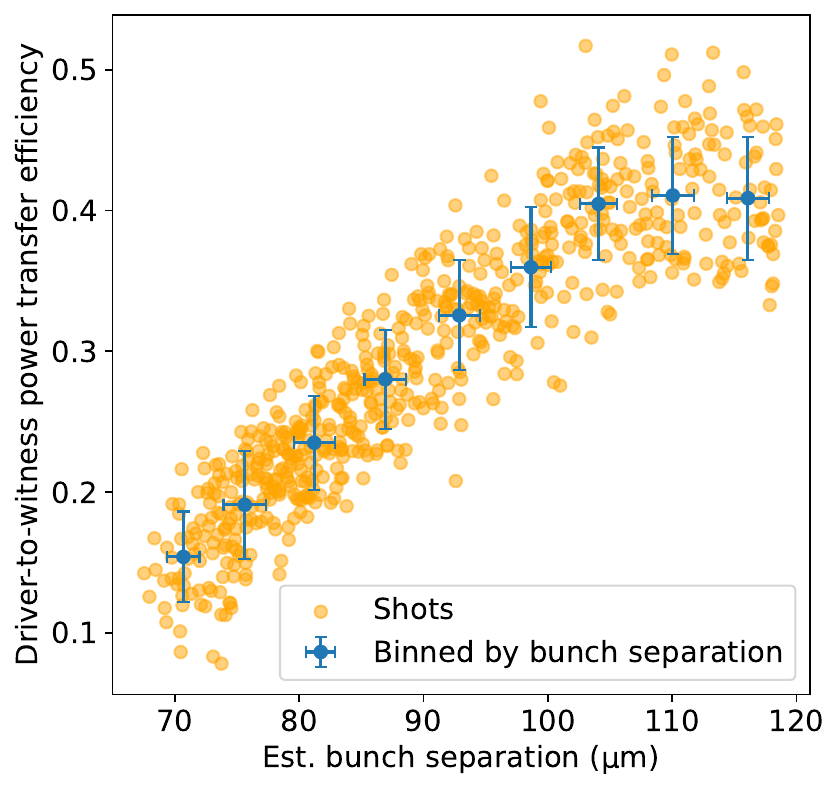}
    \caption{The correlation between bunch separation, found from converting the bunch length monitor signal using the linear fit previously extracted, and the driver-to-trailing power transfer efficiency calculated from the spectrometer screen.}
    \label{Fig3}
\end{figure}

Changing the bunch separation shifts the accelerating phase that the trailing bunch experiences within the plasma wake.  This is used to directly probe the power-transfer efficiency from the driver to the trailing bunch. We calculate the power-transfer efficiency $\eta_{p}$ using 
\begin{equation}
    \eta_{p} = \frac{\int (E_t-E_{0}) dQ_{t}}{\int (E_d-E_{0}) dQ_{d}},
\end{equation}
where $dQ_{t,d}$ and $E_{t/d}$ are the charge and per-particle energy (in GeV) of slices in the energy spectrum for the trailing and driving bunch respectively, and the initial energy $E_0$ of both the driving and trailing bunches is assumed to be 10~GeV. This model does not account for charge loss, which, in the case where charge would be lost from the driver or the witness, may result in a somewhat over- or under-estimated efficiency, respectively.
Figure~\ref{Fig3} confirms a strong correlation between bunch separation and efficiency. 

Extracting the oscillations of the trailing bunch is more challenging, as there are both horizontal and vertical kicks to the beam in the plasma; vertical kicks can change the apparent energy of the slice. Additionally, slices can have separate longitudinal positions within the bunch, but share the same energy when accelerated (due to beam loading). For both of the above reasons, the bunch can appear completely flat on the spectrometer screen at a given energy, making it difficult to trace a beam oscillation across energies with a Gaussian fit. Therefore, after converting our spectrometer images to angle--energy distributions, we find that the most robust quantification of horizontal kicks is to apply a threshold on each side of the signal region for each energy slice and obtain a $x'$ trace along the energy of the bunch.

\section{Results}
\begin{figure}[h]
    \centering
    \includegraphics[width = 1\linewidth]{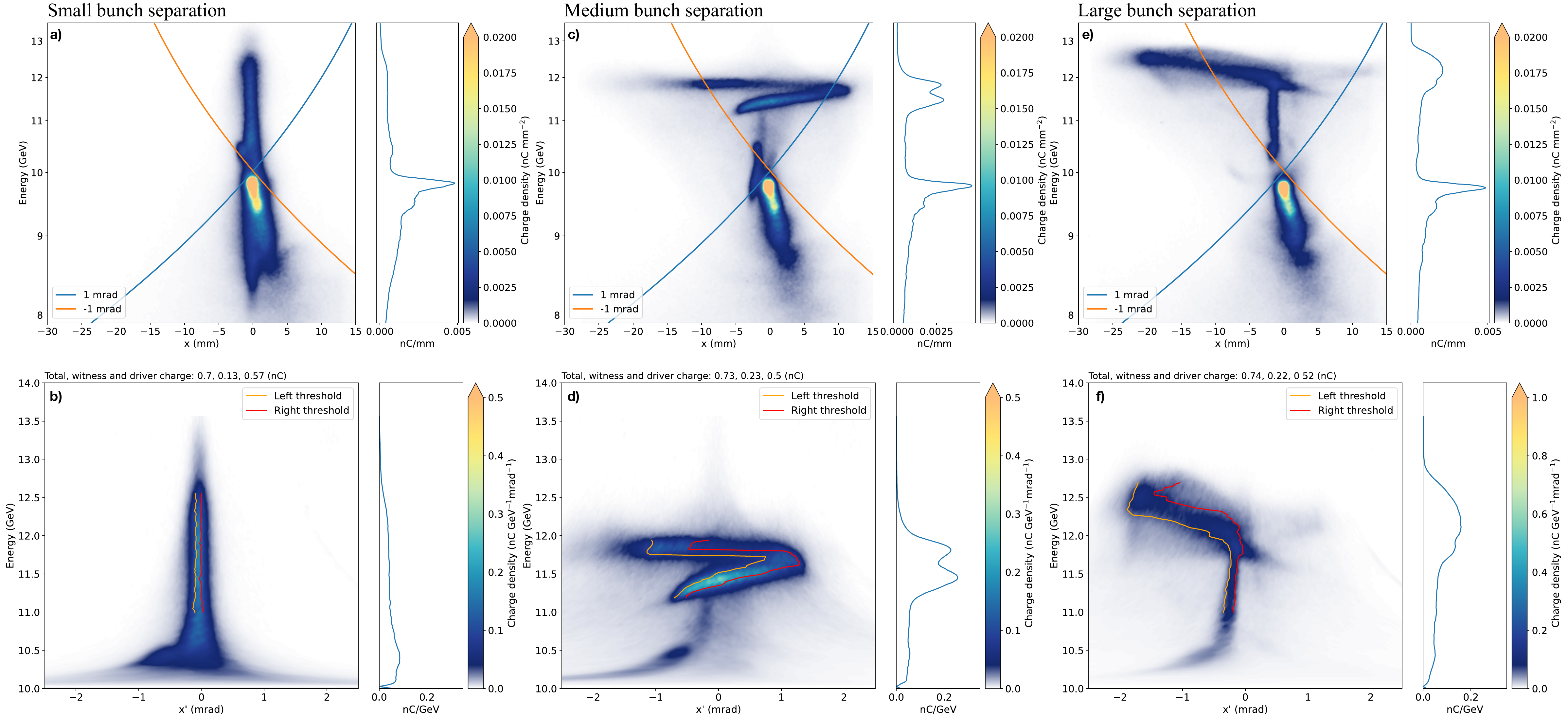}
    \caption{Three spectrometer shots (a,c,e) with lines drawn to indicate the chromatic expansion of the beam due to imaging effects for bunch slices with initial divergences of -1 and 1 mrad. The $x'$--$E$ distributions (b,d,f) are shown after conversion using the known lattice parameters. Bunch separation and efficiency are estimated to be around 71 microns and 0.17 (a,b), 95 microns and 0.36 (c,d) and 112 microns and 0.43 (e,f).}
    \label{Fig4}
\end{figure}

Figure~\ref{Fig4} shows three separate shots with increasing bunch separation, probing progressively further towards the back of the plasma cavity. 
\\
The top figure in each case is the standard spectrometer image with lines drawn on that indicate the beam expansion on the spectrometer due to the propagation of its divergence through the magnetic lattice for a given initial angle (i.e., -1 and 1 mrad). The imaging energy is 10 GeV. The bottom figures show the converted $x'$--$E$ distributions. The left (orange) and right (red) contours of the angular profile are extracted by selecting a minimum threshold value for the charge density and are used in the following data summary. 

The first shot (a,b) has a bunch separation of approximately 71 microns, calculated using the BC14 bunch length monitor value, and an efficiency extracted from the spectrometer of 0.17.
The driving bunch is stable and sees up to a few GeV of deceleration. The trailing bunch contains roughly one-fifth of the driving charge calculated from the screen, assuming all the accelerated charge is the trailing bunch and all the decelerated charge is the driving bunch. There is a significant amount of charge at 10--12 GeV, indicating a long trailing bunch that samples a region of the accelerating field. The witness bunch shows no significant transverse structure or kicks and appears unperturbed in its angular distribution.

The second shot (c,d) has a larger bunch separation of approximately 95 microns and an efficiency of approximately 0.36. The driving bunch appears similar on the spectrometer to before. The trailing bunch contains roughly one-third of the driving charge calculated from the screen. The trailing bunch has a similar energy distribution to the previous shot, although with more of its charge located at around 11 to 12 GeV, as opposed to a uniform charge density from 10 to 12 GeV. From around 11 GeV, where the angular distribution begins to dominate the transverse position on the screen due to a large $m_{12}$ (transfer matrix element), the shot has a clear oscillation in $x'$ along the energy axis. The width of this oscillation increases as we move to higher energies in the converted $x'$--$E$ distribution. The threshold filter on the converted distribution indicates an oscillation increasing in amplitude along the energy axis. The maximum transverse kick of the trailing bunch is a little over 1 mrad. 

The third shot (e,f) has an even larger bunch separation of approximately 112 microns and an efficiency of approximately 0.43. This shot has a clear transverse structure in the accelerated witness bunch. Interestingly, the bunch shows no transverse structure from around 10 GeV up until around 12 GeV, where it blows up. The angular distribution of the bunch does not show a gradual increase in the divergence of the bunch, but rather a very sharp increase at the point of highest spectral density. Additionally, the bunch sees a large accelerating field, as a significant portion of the accelerated bunch sits at around 12 GeV on the spectrometer screen. The bunch is kicked close to 2 mrad around this point. 

Since the trailing bunch for the shot with the smallest bunch separation (a) has approximately 70 pC less charge than the subsequent shots (c,e), which is similar to the added charge the driver has compared to the later shots, it is likely that at 71 microns bunch separation (a), some of the driving bunch is located in the accelerated phase.

\begin{figure}[h!]
    \centering
    \includegraphics[width = 0.7\linewidth]{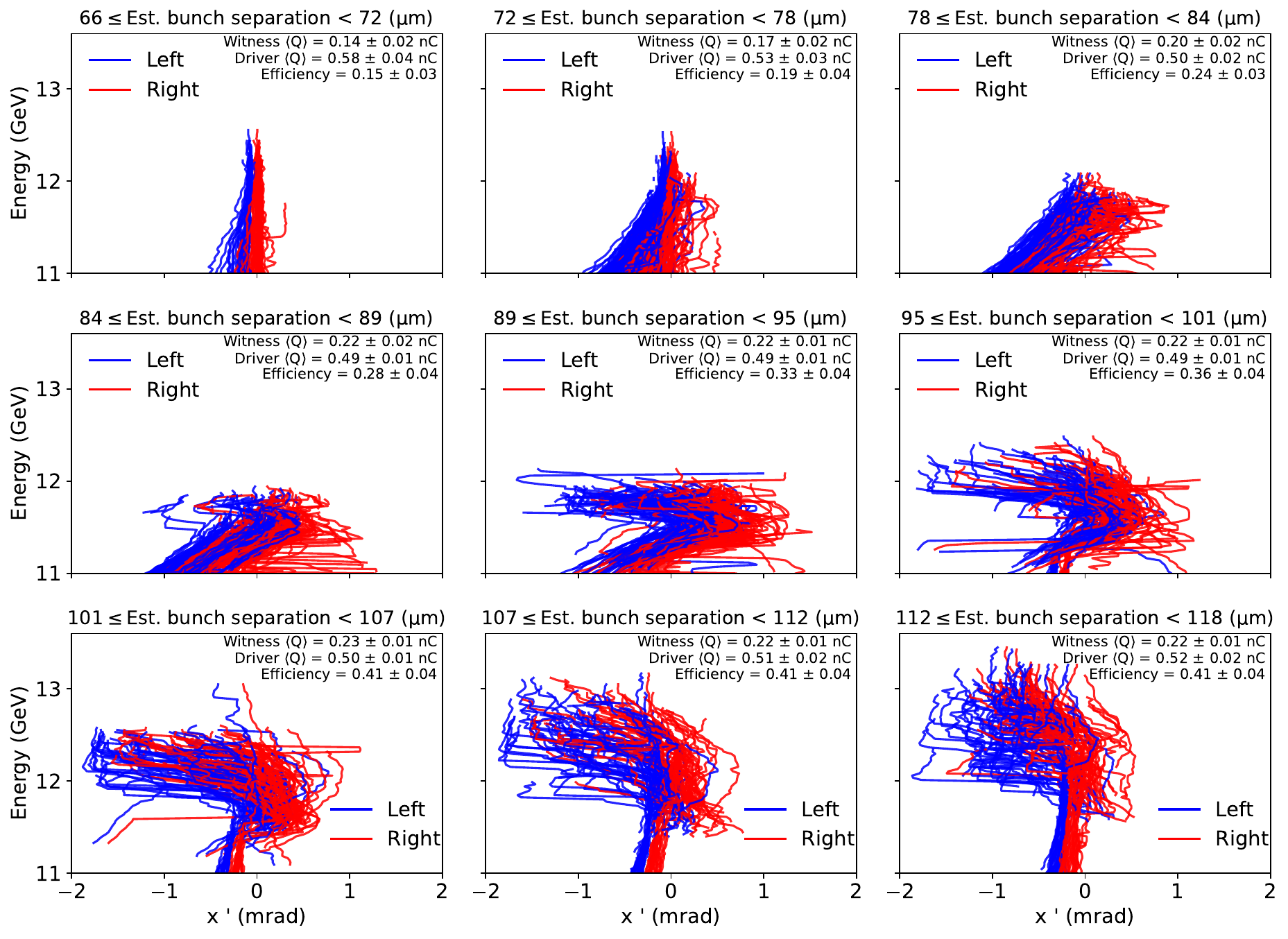}
    \caption{Left and right threshold value lines drawn for each shot in the scan of the L2 phase. Mean charges and efficiency calculated from the spectrometer screen are annotated on each window of the figure for each estimated bunch separation interval, along with their standard deviation.}
    \label{Fig5}
\end{figure}

Figure~\ref{Fig5} shows the left and right contour for each shot in the dataset, binned by estimated bunch separation value. To ensure that we are mainly imaging the angular distribution of the beam, the energy axis is cut below 11 GeV, where the angular contribution to the position on the spectrometer screen is roughly twice that of the initial position. We observe clear, systematic changes in the bunch's evolution throughout the scan. Initially, at low bunch separation values between 66 and 72 microns, there is no significant oscillation or curve in the lines, and the efficiency is very low. Then, in the next few steps, the lines begin to curve before a fully formed oscillation is observed around bunch separation values of 89 microns and above. Moving past this point, the lines begin to appear straight at lower energies up until a point where there is a large, consistent kick in $x'$. Here, there are several shots in each step approaching a maximum angle of 2 mrad. As bunch separation increases, the energy at which strong oscillations occur increases. 

At low bunch separation, some of the trailing bunch is likely in the decelerating phase of the wakefield, as the charge measured from the spectrometer screen is lower than in subsequent steps, which are fairly consistent around 0.22 nC. This is further indicated by the driver having more charge in the earlier steps. The total amount of charge, however, is roughly 0.7 nC across all steps.

\begin{figure}
    \centering
    \includegraphics[width = 0.7\linewidth]{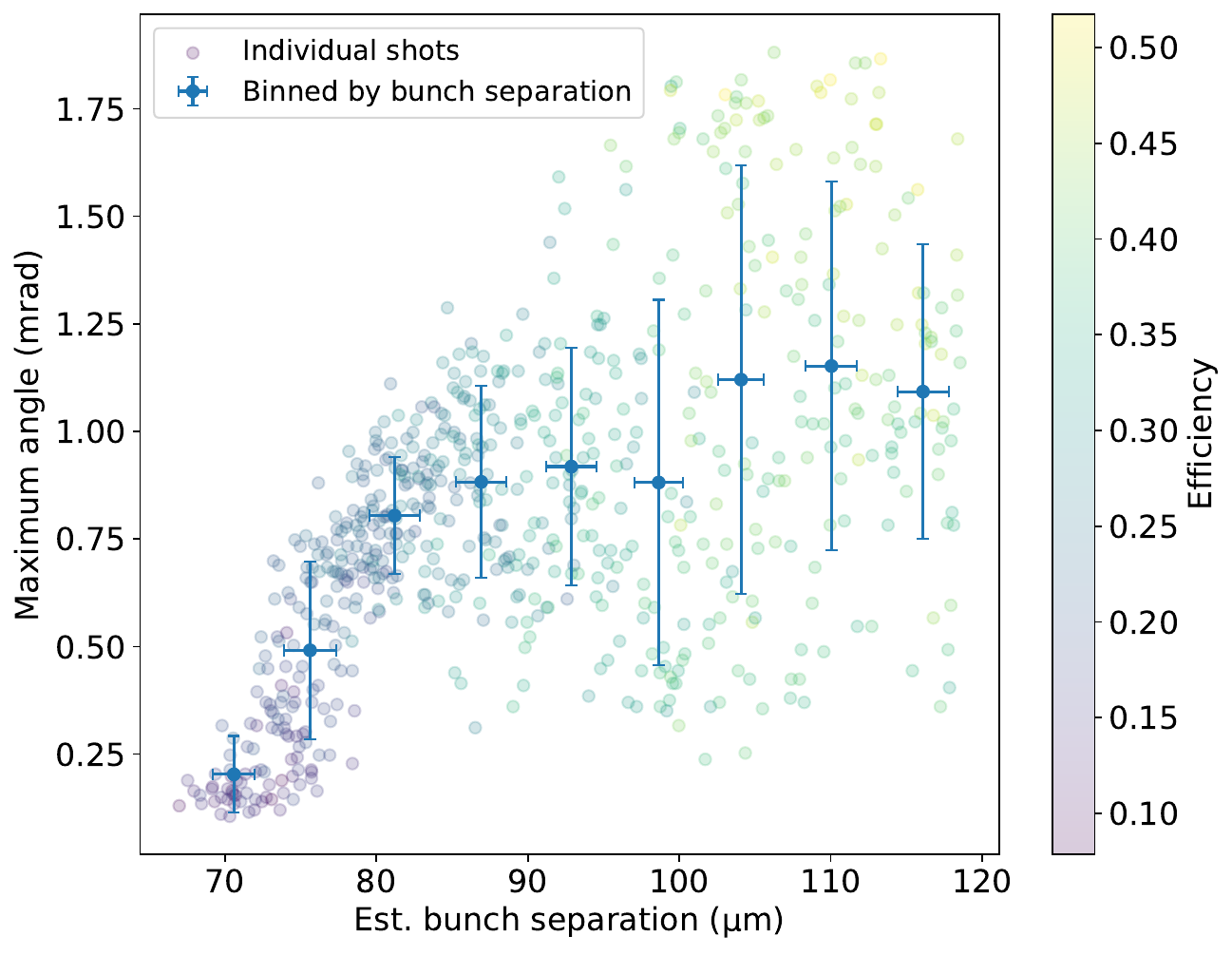}
    \caption{The maximum absolute angle reached by parts of the bunch using a left and right threshold filter on each shot in the dataset is shown against bunch separation found from applying the linear calibration to the bunch length monitor signal.}
    \label{Fig6}
\end{figure}

The maximum angle recorded for each shot in the dataset is shown against bunch separation and efficiency in Fig.~\ref{Fig6}. There is a clear correlation between the maximum angle and bunch separation, and thus efficiency. Initially, the parameters correlate well, and the maximum angle increases rapidly with bunch separation. However, at larger bunch separation and efficiency, there is a large spread in the maximum angles, ranging from around 0.25 mrad to above 1.75 mrad.

\section{Simulation}

The results are compared with simulations in order to examine the BBU instability dependences with all parameters known. 

\begin{figure}[ht]
    \centering
    \includegraphics[width = 0.6\linewidth]{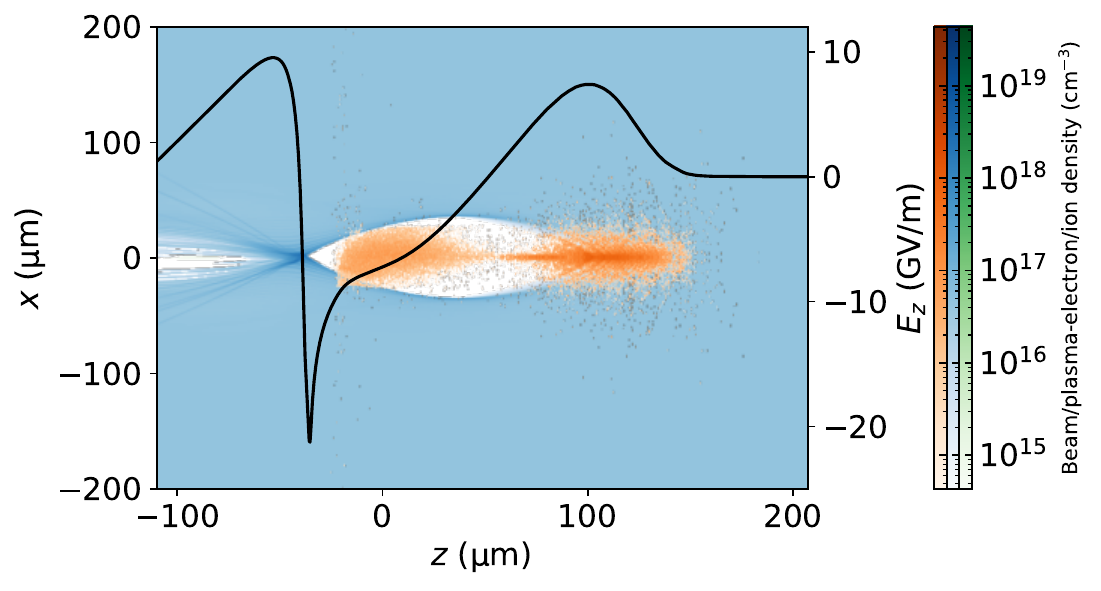}
    \caption{HiPACE++ simulation of the experiment, including transverse instability. Beam, plasma, and ion density are shown for a time step in the middle of the plasma. The plasma background density is subtracted from the ion density (no ion over-density is observed). Here, the bunch separation is 111 $\upmu\mathrm{m}$.}
    \label{Fig7}
\end{figure}

We simulate two Gaussian bunches through the FACET-II plasma stage, including ramps, transported through the experimental beamline onto the FACET-II spectrometer screen. The plasma interaction is modelled using HiPACE++ \cite{HiPACE}, whereas the spectrometer is modelled using ImpactX \cite{IMPACTX}.
The transverse Twiss parameters at the waist are extracted from a scan of the magnet strengths without plasma. Unfortunately, some shots in these datasets were saturated on the diagnostic screen, hence there is a potential for small errors in these input parameters. 
We assume 20 micron length bunches, which are on the order of those used at FACET-II, and give an energy distribution on the simulated spectrometer that is similar to what we see in experiment. An offset for the trailing bunch is introduced in $x$ with a corresponding jitter. We use the ABEL framework \cite{ABEL} to couple the different codes used for the plasma interaction and the spectrometer, enabling a start-to-end simulation of the experiment. A wake plot of a shot at the middle point in the plasma in the simulation with a bunch separation of 111 $\upmu\mathrm{m}$ is shown in Fig.~\ref{Fig7}. 

The trailing bunch is not matched to the plasma and therefore undergoes envelope oscillations during acceleration. Across the scan, the field is nearly flat at the beginning of the plasma cell. However, during acceleration, the field deviates significantly from being flattened. Multiple effects, such as drive-beam energy loss \cite{energyspreadhosing} and charge loss in the trailing bunch, contribute to changes in field loading. 

\begin{figure}[h]
    \centering
    \includegraphics[width = 0.6\linewidth]{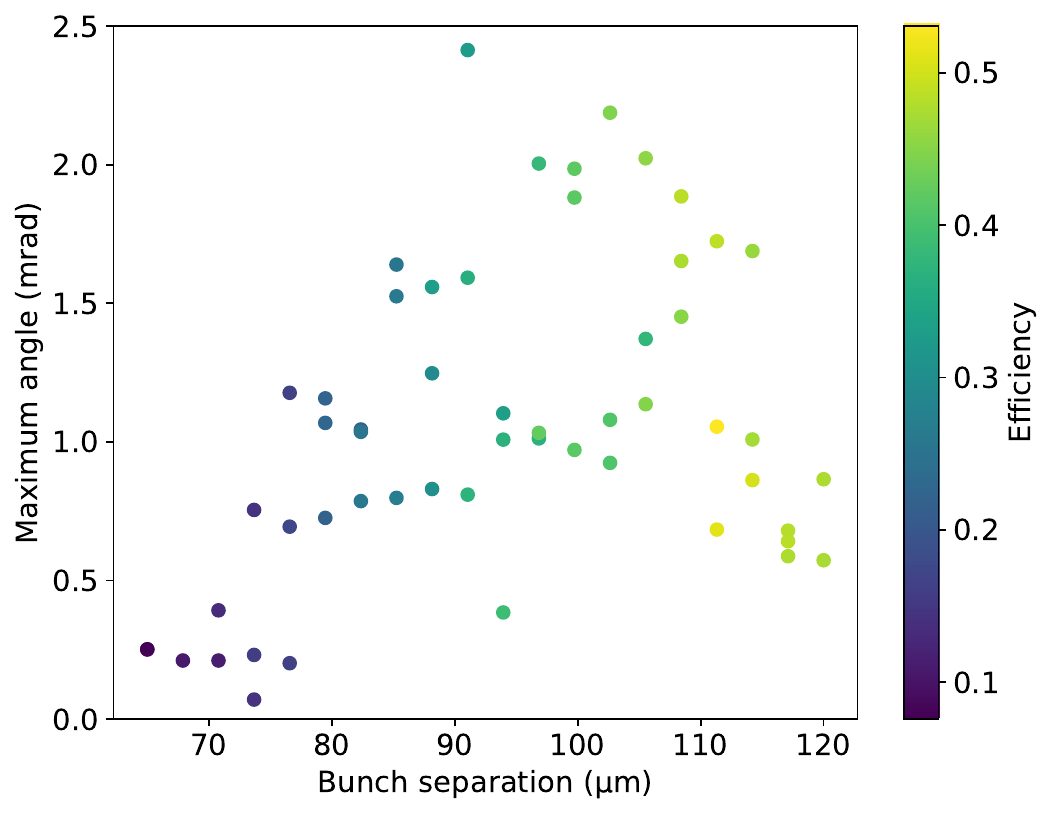}
    \caption{Simulated maximum angle versus bunch separation from HiPACE++. The maximum absolute angle reached by parts of the bunch using a left and right threshold filter, as well as the efficiency, is calculated for each shot using the simulated spectrometer screen.}
    \label{Fig8}
\end{figure}

In Fig.~\ref{Fig8}, the maximum angle of the threshold lines drawn on the simulated data is shown against bunch separation, with efficiency extracted from the spectrometer shown on the color bar. 
At low bunch separation and efficiency, the maximum angles for each shot are small. As bunch separation and, consequently, efficiency increase, the maximum angles extracted increase rapidly, similar to the behaviour observed in the previously shown experimental data in Fig.~\ref{Fig6}. Also present in both the experimental and simulated data is the wide range of maximum angles extracted at high bunch separation and efficiency. At around 100 micron bunch separation, there are shots with maximum angle ranging from under 1 mrad to over 2 mrad. 

\begin{figure}[h]
    \centering
    \includegraphics[width = 0.6\linewidth]{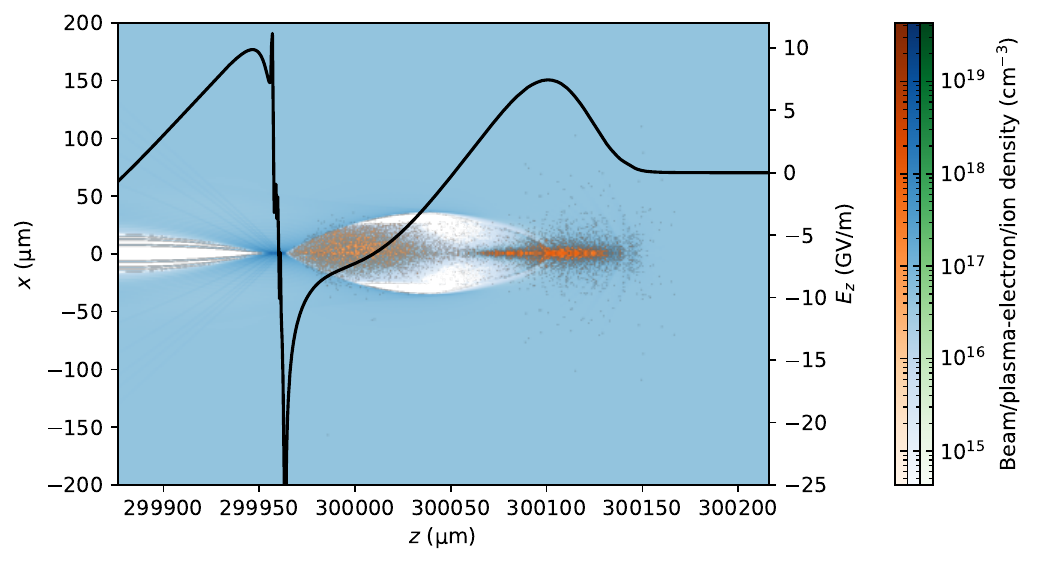}
    \caption{WakeT simulation of the experiment, not including transverse instability. Beam, plasma, and ion density are shown for a time step in the middle of the plasma. The plasma background density is subtracted from the ion density (no ion over-density is observed). Here, the bunch separation is 111 $\upmu\mathrm{m}$.}
    \label{Fig9}
\end{figure}

\begin{figure}[ht]
    \centering
    \includegraphics[width = 0.6\linewidth]{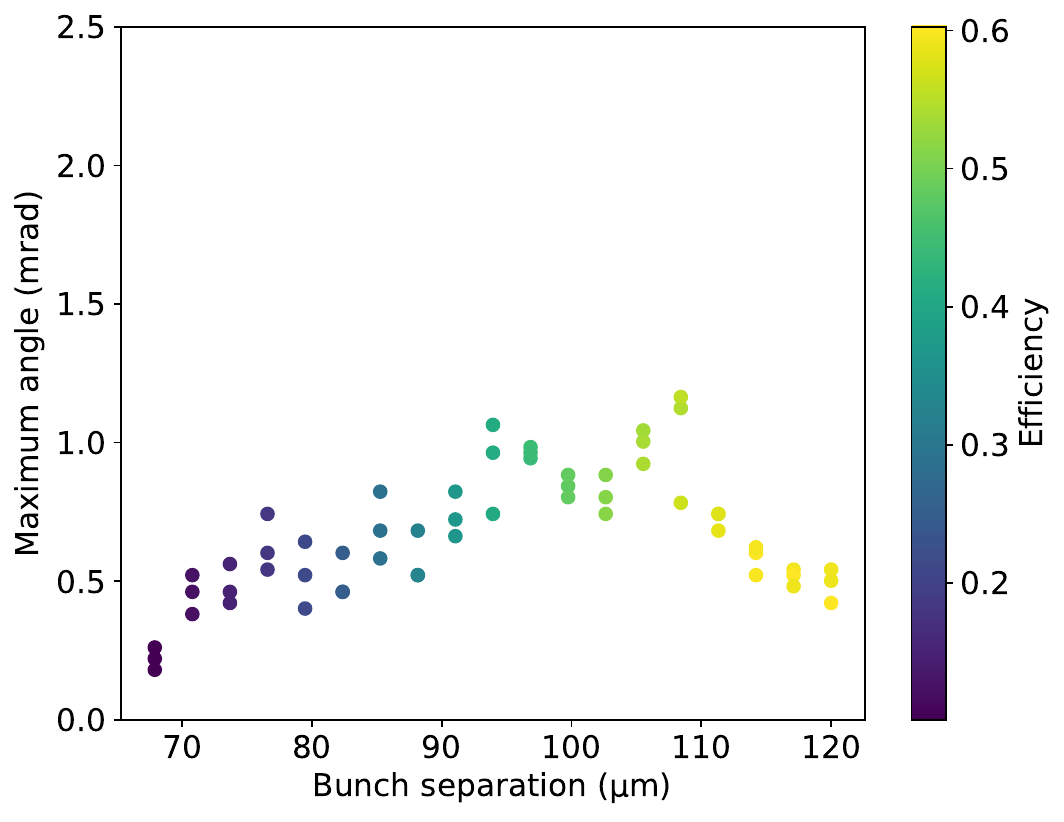}
    \caption{Simulated maximum angle versus bunch separation from WakeT. Again, the maximum absolute angle reached by parts of the bunch using a left and right threshold filter, as well as the efficiency, is calculated for each shot using the simulated spectrometer screen.}
    \label{Fig10}
\end{figure}

Simulations let us isolate the effect of the instability from the remaining, complex beam dynamics in the plasma. We have redone the simulations with identical input parameters with Wake-T \cite{WakeT}, an analytical particle-tracking code with cylindrically-symmetric fields. Because of the symmetry, transverse instabilities such as the BBU instability are absent. A wake plot of the same shot as shown in Fig.~\ref{Fig7}, simulated without instabilities, is shown in Fig.~\ref{Fig9}. In Fig.~\ref{Fig10}, the maximum angle extracted from the contour lines of the Wake-T simulations is shown against bunch separation. The maximum angles are closer to 1 mrad, compared with around 2 mrad for the HiPACE++ results; a significant decrease indicating the importance of the transverse instability. The spread of maximum angle reached is much smaller as well, from around 0.5 mrad to 1 mrad, indicating less chaotic behavior.

We observe that simulations that include the transverse instability (Fig.~\ref{Fig8}) are quantitatively closer to the experiment (Fig.~\ref{Fig6}) than those without instability (Fig.~\ref{Fig10}).

\section{Discussion}

At small bunch separation, we observe only small angles, and as bunch separation and consequently efficiency increase, the maximum angles increase. At higher energies, we see a large and sudden increase in the $x'$ value reached by these lines. Moreover, this behaviour is very consistent across the dataset. 

This may be explained by an effect that opposes the build-up of the instability along the tail of the trailing bunch: Balakin-Novokhatsky-Smirnov damping (BNS damping) \cite{Mehrlingchirp, lebedev,Parametricmapping}. If the bunch is sufficiently long or has a low beam current, the field is not flattened by beam loading, and we have a chirp in the field which induces phase mixing in the betatron oscillations of the electrons, and the resonance of the instability is disrupted. At very high bunch separation, a significant amount of charge is located at higher energies, hence the instability is suppressed at lower energy slices where the beam does not load the wake. Coupled with field changes during acceleration that induce an energy spread in the trailing bunch, this can explain why there are fewer transverse kicks at lower energies in Fig.~\ref{Fig5}. In Fig.~\ref{Fig4}(e,f), where the bunch separation is estimated to be roughly 112 microns, there is a large energy spread at lower energies where there are little transverse kicks. The point at which it rapidly grows in $x'$, most of the charge is situated in a smaller range of energies, hence it is likely that the accelerating field flattened out during acceleration (optimally beam loaded), and BNS damping does not occur in this part of the spectrum. In Fig.~\ref{Fig5}, the point at which the transverse kicks become significant moves towards higher energies, visible from bunch separation values of 95 to 118 microns.

In short, the behaviour of the transverse kicks shown in Fig.~\ref{Fig6} and Fig.~\ref{Fig8} could be understood by considering the instability--efficiency relation, beam loading, and BNS damping. At low bunch separation and therefore efficiency, the instability is not large enough to build up past small to medium transverse kicks. Hence, the maximum angles at low bunch separation are smaller and increase systematically as the bunch separation increases. However, at large bunch separations and efficiencies, the instability already has a large seed, to the point that BNS damping cannot significantly mitigate the amplitude growth. Coupled with jitter in the initial transverse offset of the trailing bunch, the result is a violently unstable process in which a small deviation in the initial setup leads to a very large difference in the angle at the plasma exit. Finally, at high bunch separation, as the trailing bunch moves farther behind the driving bunch, the wake radius decreases. Hence, as the trailing bunch oscillates, some electrons are ejected from the wake. This can also produce small angles, as the charge in slices with large transverse kicks is sometimes lost.

\section{Conclusions}



We have extracted the maximum angle of transverse oscillations by analysis of the transverse contours of single shots. At lower bunch separations, the FACET-II data show a clear correlation between maximum angle, the bunch separation, and thus the efficiency, consistent with the build-up of an instability. At larger bunch separations, the correlation is less strong.  Furthermore, the onset of transverse oscillations occurs towards higher energies for larger bunch separation.

Numerical simulations of the experiment, with and without the transverse instability enabled, indicate that the data are similar to simulations with the instability present. The simulations indicate that BNS damping due to the trailing bunch energy spread, because of under-loading of the accelerating field, may explain the delay in the onset of the instability at higher separation.




For both the experimental data and the simulated data, the maximum angles are reached around 100 to 110 micron bunch spacing. This is likely because the instability is present, but small enough that, with a large wake radius, the transverse kicks are contained within the wake. At very large bunch spacing, the large instability, coupled with a small wake radius, limits the transverse kicks achievable before significant charge loss. 

The methods outlined in this paper can be used to extract the angle--energy distribution of the bunch after the plasma and correlate the angles at each energy slice on the spectrometer screen with efficiency. Future work on this subject should build on these methods to quantify more precisely the instability's effect on the trailing bunch and the relation between efficiency and instability \cite{lebedev,Parametricmapping}.

The experiment can be further improved by high-resolution transverse diagnostics, capable of detecting a wider range of transverse kicks. Additionally, as indicated by the simulations (Fig.~\ref{Fig7} and Fig.~\ref{Fig9}), the plasma blow-out radius is not much larger than the beam, leaving little transverse space for the instability to grow. A drive beam providing a stronger blow-out is therefore desired. These improvements will make it possible to generate and measure larger transverse kicks from the instability and the resonant behaviour, making it easier to precisely quantify the instability. Mapping and understanding of the wakefields induced in the plasma is also essential for precise, quantitative measurements of instability and efficiency. Lastly, approximate parameters were used as inputs for the simulations. Fully characterizing the beam's longitudinal and transverse properties with high precision would also be a step towards a precise one-to-one comparison with experimental data. For this purpose, diagnostics to precisely estimate the offsets of both the driving and trailing bunches, on a shot-to-shot basis, would be crucial.

\section*{Acknowledgments}
This work was supported by the Research Council of Norway (NFR Grant No.~313770) and the European Research Council (ERC Grant No.~101116161). We acknowledge Sigma2 - the National Infrastructure for High-Performance Computing and Data Storage in Norway for awarding this project access to the LUMI supercomputer, owned by the EuroHPC Joint Undertaking, hosted by CSC (Finland) and the LUMI consortium. FACET-II is supported in part by the U.S. Department of Energy under contract number DE-AC02-76SF00515.

\end{document}